\documentclass[10pt,twoside]{article}
\usepackage{amsmath}    
\usepackage{amsfonts}  
\usepackage{amssymb}
\usepackage{graphicx}   
\usepackage{array} 
\usepackage{lipsum}

\usepackage{float}   
\usepackage[T1]{fontenc}
\usepackage[latin1]{inputenc}     
\usepackage[title]{appendix}
\usepackage{amsmath}    
\usepackage{amsfonts}  
\usepackage{amssymb}
\usepackage{graphicx}   

\usepackage[latin1]{inputenc}     
\usepackage[T1]{fontenc}

\usepackage{float}   
\usepackage{tikz}

\usepackage[a4paper]{geometry} 
\usepackage{color}   
\usepackage{soul}    
\definecolor{lightblue}{rgb}{0.8,0.9,1} 
\definecolor{lightred}{rgb}{1,0.5,0.4} 
\definecolor{lightgreen}{rgb}{0.4,1,0.4} 

\usepackage{ulem} 
\author{Michel Gondran\\\textit{Académie Européenne Interdisciplinaire des Sciences, Paris, France}\\\texttt{michel.gondran@polytechnique.org}\\~\\
Alexandre Gondran\\\textit{ENAC, Toulouse University, Toulouse,France}\\\texttt{alexandre.gondran@recherche.enac.fr}}

\title{A pseudo-random and non-point Nelson-style process}

\newtheorem{Theoreme}{THEOREM}
\newtheorem{Definition}{Definition}

\newtheorem{Proposition}{Proposition}
\newtheorem{Remarque}{Remark}

\graphicspath{
{/home/alexandre/Documents/mq/images/}{/home/alexandre/Documents/mq/images/portraits/}{/home/alexandre/Documents/mq/images/sternGerlach/}{/home/alexandre/Documents/mq/images/young/}{/home/alexandre/Documents/mq/images/epr/}
{/home/gondran/Documents/mq/images/}{/home/gondran/Documents/mq/images/portraits/}{/home/gondran/Documents/mq/images/sternGerlach/}{/home/gondran/Documents/mq/images/young/}{/home/gondran/Documents/mq/images/epr/}}

\date{June, 2024}

\begin{document}
\maketitle
\vskip 1cm

\begin{abstract}
We take up the idea of Nelson's stochastic processes, the aim of which was to deduce Schrödinger's equation. We make two major changes here. The first one is to consider deterministic processes which are pseudo-random but which have the same characteristics as Nelson's stochastic processes. The second is to consider an extended particle and to represent it by a set of interacting vibrating points.

In a first step, we represent the particle and its evolution by four points that define the structure of a small elastic string that vibrates, alternating at each period a creative process followed by a process of annihilation. We then show how Heisenberg's spin and relations of uncertainty emerge from this extended particle.

In a second step, we show how a complex action associated with this extended particle verifies, from a generalized least action principle, a complex second-order Hamilton-Jacobi equation. We then deduce that the wave function, accepting this complex action as a phase, is the solution to a Schrödinger equation and that the center of gravity of this extended particle follows the trajectories of de Broglie-Bohm's interpretation.

This extended particle model is built on two new mathematical concepts that we have introduced: complex analytical mechanics on functions with complex values \cite{Gondran2001a,Gondran2001c,Gondran2003} and periodic deterministic processes \cite{Gondran2001a,Gondran2004a}. 

In conclusion, we show that this particle model and its associated wave function are compatible with the quantum mechanical interpretation of the double-scale theory we recently proposed \cite{Gondran2021}. 

{\it 
~\\\textbf{Résumé} -  
Nous reprenons l'idée des processus stochastiques de Nelson, dont le but était de déduire l'équation de Schrödinger. Nous apportons ici deux changements majeurs. Le  premier est de considérer des processus déterministes pseudo-aléatoires mais qui ont les mêmes caractéristiques que les processus stochastiques de Nelson. Le second est de considérer une particule étendue et de la représenter par un ensemble de points vibrants en interaction.

Dans un premier temps, nous représentons la particule et son évolution par quatre points qui ont la structure d'une petite corde élastique qui vibre en alternant à chaque période un processus de création suivi d'un processus d'annihilation. Nous montrons ensuite comment le spin  et les relations d'incertitude d'Heisenberg émergent de cette particule étendue.

Dans un deuxième temps, nous montrons comment une action complexe associée à cette particule étendue  vérifie, à partir d'un principe de moindre action généralisé, une équation de Hamilton-Jacobi complexe du second ordre. 

Nous déduisons ensuite que la fonction d'onde, acceptant cette action complexe comme phase, est la solution d'une équation de Schrödinger et que le centre de gravité de cette particule étendue suit les trajectoires de l'onde pilote de de Broglie-Bohm.

Ce modèle de particule étendue est construit sur deux nouveaux concepts que nous avons introduit : une mécanique analytique complexe à valeurs complexes et des processus déterministe périodique.

En conclusion, nous montrons que ce modèle de particule et sa fonction d'onde associée sont compatibles avec la théorie de la double solution de de Broglie et la théorie de la double échelle que nous avons proposée récemment.}

\end{abstract}

\graphicspath{
{/home/alexandre/Documents/mq/images/}{/home/alexandre/Documents/mq/images/portraits/}{/home/alexandre/Documents/mq/images/sternGerlach/}{/home/alexandre/Documents/mq/images/young/}{/home/alexandre/Documents/mq/images/epr/}
{/home/gondran/Documents/mq/images/}{/home/gondran/Documents/mq/images/portraits/}{/home/gondran/Documents/mq/images/sternGerlach/}{/home/gondran/Documents/mq/images/young/}{/home/gondran/Documents/mq/images/epr/}}


\section{Introduction}

Stochastic mechanics, developed and popularized by Nelson \cite{Nelson1966, Nelson1967, Nelson1985}, is an interpretation of quantum mechanics using diffusion theory. It presents, as well as its various variants \cite{Nelson2012}, an alternative to the foundations of quantum mechanics. Nelson's program was to derive the wave function and the Schrödinger equation from a double scattering process.
In one of his final papers \cite{Nelson2012} in 2012, Nelson, after taking stock of the successes and failures of stochastic mechanics, concludes his article as follows:
 
"\textit{How can a theory to be so right and yet so wrong? The most natural explanation is that stochastic mechanics is an approximation to a correct theory of quantum mechanics as emergent. But what is the correct theory?}"

The aim of this paper is to present an alternative to stochastic mechanics which can allow us to approach the correct theory by realising part of Nelson's program: demonstrating Schrödinger's equation and proposing realistic trajectories. To do this, we take up Nelson's approach with two major changes. The first is to consider pseudo-random deterministic processes with the same statistical characteristics as Nelson's stochastic processes. The second change is to consider processes corresponding to different points of an extended particle. For pedagogical reasons, we first consider the case in dimension one before extending it to dimension two. 

To build this model, we generalize to quantum mechanics the approach used in classical mechanics to define Hamilton-Jacobi's action from the principle of least action.

The outline of the paper is as follows. In section 2, we briefly recall how the principle of least action makes it possible to obtain the Hamilton-Jacobi equation. In section 3, we describe in dimension 1 the evolution of an extended particle represented by two points that simulate the structure of a small elastic cord that vibrates, alternating at each period a sort of creative process followed by a process of annihilation. In section 4, we generalize this model in dimension two with an extended particle represented by four points. In section 5, we show how Heisenberg's spin and uncertainty relations emerge from this extended particle in dimension 2. In section 6, we show how a complex action associated with this extended particle verifies, from a generalized principle of least action, a complex second order Hamilton-Jacobi equation. Then in section 7, we show that the wave function, admitting this complex action as a phase, is the solution to a Schrödinger equation and that the center of gravity of this extended particle follows the trajectories of de Broglie-Bohm's interpretation.

This model is built on two new mathematical concepts: the complex analytical mechanics on complex value functions that we introduced 
\cite{Gondran2001a,Gondran2001c,Gondran2003}
and the periodic deterministic processes that
we have developed
\cite{Gondran2001a,Gondran2004a}.

\section{The Principle of Least Action and the Hamilton-Jacobi Equation}

Let us begin by recalling how the principle of least action applies in classical mechanics in order to deduce Hamilton-Jacobi's action and demonstrate how we can deduce that everything happens as if this action piloted the classical particle.

\subsection{The Hamilton-Jacobi action "pilots" the classical particle}

Let us consider in classical mechanics the center-of-mass of a rigid particle. For any time step $\varepsilon > 0$, the evolution of this center-of-mass, at time $s= k \varepsilon$ ($s \in [\varepsilon, N \varepsilon]$), is defined by the equations 
\begin{equation}\label{eq:evc}
\textbf{x}_\varepsilon (k \varepsilon) = \textbf{x}_\varepsilon ((k-1)\varepsilon) + \textbf{u }(k \varepsilon) \varepsilon~~~~with~~~~ \textbf{x}_\varepsilon(0)=\textbf{x}(0)
\end{equation}
where $\textbf{u}:\mathbb{R}^+\to\mathbb{R}^n$ (with $n=1,2$ or $3$) is a continuously differentiable function and $\textbf{x}(0) \in \mathbb{R}^{n}$ a given initial position. 
We also set at the initial instant, an initial action $S^0(\textbf{x})$, a function of
$\mathbb{R}^{n}$ in $\mathbb{R}$. We show that this initial action corresponds to the initial velocity field $\textbf{v}^0(\textbf{x})= \nabla S^0(\textbf{x})/m $.

The Hamilton-Jacobi action $S(\mathbf{x},t)$ is then the function~:
\begin{equation}\label{eq:defactionHJb}
S(\mathbf{x},t)=\underset{\textbf{x}_0;\mathbf{u}(.)
}{\min
}\left\{ S^{0}\left( \mathbf{x}_{0}\right) +\int_{0}^{t}L(\textbf{x}(s),%
\mathbf{u}(s),s)ds\right\}
\end{equation}
where the minimum (or more generally the extremum) of (\ref{eq:defactionHJb}) is taken over all the trajectories with starting points $\textbf{x}_0$ and the end point $\textbf{x}$ at $t= N\varepsilon$ as well over velocities along the entire trajectory
$\mathbf{u}(s)$, with $s\in$ $\left[ 0,t\right]$ and $L(\textbf{x}, \textbf{u}, t)$ is the Lagrangian.
Hamilto n-Jacobi's discreet action is written:
\begin{equation}\label{eq:defactionHJc}
S_\varepsilon(\mathbf{x},t=N\varepsilon)=\underset{\textbf{x}_0;\mathbf{u}(.)}{\min
}\left\{ S^{0}\left( \mathbf{x}_{0}\right) +\sum_{k=1}^{k=N}L(\textbf{x}(k \varepsilon),%
\mathbf{u}(k \varepsilon),k \varepsilon)\varepsilon\right\}
\end{equation}
where the minimum (or more generally the extremum) of (\ref{eq:defactionHJc}) is taken over all the
discrete trajectories (\ref{eq:evc}) with starting points $\textbf{x}_0 $ and the end point $\textbf{x}$ at $t$ and over the velocities $\mathbf{u}(s)$, with $s =k \varepsilon, k\in$ $\left[1,N \right]$.

Hamilton-Jacobi's discrete action
$S_\varepsilon(\mathbf{x},t)$ then verifies, between the instants t-$\varepsilon $ and t, the optimality equation:

\begin{equation}\label{eq:defactionlocaleb}
S_\varepsilon(\mathbf{x},t)=\underset{\mathbf{u}\left(t\right)}{\min }\left\{ S_\varepsilon(\mathbf{x}-\textbf{u}(t)
\varepsilon,t- \varepsilon) + L(\textbf{x},
\mathbf{u}(t),t)\varepsilon\right\}.
\end{equation}

\begin{Remarque}\label{r:opt}- It is the optimality equation (\ref{eq:defactionlocaleb}), and not equations (\ref{eq:defactionHJb}) and (\ref{eq:defactionHJc}) that corresponds to the application of the principle of least action for the Hamilton-Jacobi action. It is therefore this optimality equation that we generalize in section 5. The locality of this optimality equation explains why the equations (\ref{eq:defactionHJb}) and (\ref{eq:defactionHJc}) are not always minima, but only extrema.
\end{Remarque}

Assuming $S_\varepsilon$ differentiable in $\textbf{x}$ and $t$, $L$ differentiable in $\textbf{x}$, $\textbf{u}$ and $t$, and that
$\textbf{u}(t)$ continues, the equation (\ref{eq:defactionlocaleb})
becomes:

\begin{equation}\label{eq:eqoptHJ}
0=\underset{\mathbf{u}\left( t\right) }{\min }\left\{
-\frac{\partial S_\varepsilon}{
\partial \mathbf{x}}\left( \mathbf{x} \mathbf{,}t\right)
\textbf{u}(t) \varepsilon-\frac{\partial S_\varepsilon}{\partial \mathbf{t%
}}\left( \mathbf{x} \mathbf{,}t\right)\varepsilon+L(\mathbf{x},
\mathbf{u}(t),t)\varepsilon+\circ \left( \varepsilon\right) \right\}
\end{equation}
i.e. by dividing by $\varepsilon$ and making $\varepsilon$ tends towards
$0^{+}$,
\begin{equation*}
\frac{\partial S }{\partial \mathbf{t}}\left( \mathbf{x,}t\right)
=\underset{ \mathbf{u}}{\min }\left\{
L(\mathbf{x},\mathbf{u},t)-\mathbf{u\cdot }\frac{
\partial S}{\partial \mathbf{x}}\left( \mathbf{x,}t\right) \right\}
\end{equation*}
hence the classic Hamilton-Jacobi equation:
\begin{equation*}
\frac{\partial S}{\partial t}\left( \mathbf{x,}t\right)
+H\left(\mathbf{x},\frac{
\partial S}{\partial \mathbf{x}}{\LARGE ,}t\right)=0\text{ \ }
\end{equation*}
where $H(\mathbf{x},\mathbf{p},t)$  is the Fenchel-Legendre transform
$L(\mathbf{x},\mathbf{u},t)$ in relation to $\mathbf{u}$. In the case of a non-relativistic particle in a field of
potential $V(\mathbf{x},t)$, this yields the following theorem:

\begin{Theoreme}\label{th:emergencespin}\cite{Gondran2021}- 
The Hamilton-Jacobi action $S\left( \mathbf{x,}t\right) $ is the solution to the Hamilton-Jacobi equations:
\begin{equation}\label{eq:HJ}
\frac{\partial S}{\partial t}+\frac{1}{2m}(\triangledown S)^{2}+V(\textbf{x},t)=0\text{ \ \ \ \ \ \ \ }\forall \left(
\textbf{x},t\right) \in \mathbb{R} ^{n}\times \mathbb{R}^{+}
\end{equation}
\begin{equation}\label{eq:condinitialHJ}
S(\textbf{x},0)=S^{0}(\textbf{x})\text{ \ \ \
\ \ }\forall \textbf{x}\in \mathbb{R} ^{n}
\end{equation}
and the speed of the classical non-relativistic particle is given in each point $ \left(
\mathbf{x,}t\right)$ by the velocity field:
\begin{equation}\label{eq:eqvitesse}
\mathbf{u}\left( \mathbf{x,}t\right) =\frac{\mathbf{\nabla }S\left( \mathbf{%
x,}t\right) }{m} \text{ \ \ \ \ \ \ \ }\forall \left(
\textbf{x},t\right) \in \mathbb{R} ^{n}\times \mathbb{R}^{+}
\end{equation}
\end{Theoreme}

Thus the whole evolution of a classical particle depends on the data in the $S^0(\textbf{x})$ field (initial action) and on $\textbf{x}(0)$ (initial position). The Hamilton-Jacobi action is a field and everything happens as if it pilots the classical particle $\textbf{x}(t)$.

\subsection{Generalizations  for the transition from classical to quantum}

In order to implement a similar approach to quantum physics, several generalizations are necessary. 
The first generalization is to leave, for positions and for action, the space of real numbers to move to complex numbers. This is related to Schrödinger's equation of a non-relativistic quantum particle in a field of potential $V(\textbf{x},t)$:
\begin{equation}\label{eq:Schrod1}
i \hbar \dfrac{\partial \Psi}{\partial t}= -\dfrac{\hbar^2}{2 m}\Delta \Psi +V \Psi~~~~~~~~~with~~~~~~~~\Psi(\textbf{x},0)= \Psi_0(\textbf{x})
\end{equation}
Indeed, by replacing in Schrödinger's equation the wave function $\Psi(\textbf{x},t) $ by its complex phase $S(\textbf{x},t)$ thanks to the change of variable $\Psi(\textbf{x},t)=e^{i\frac{S(\textbf{x},t)}{\hbar}} $, we obtain the complex Hamilton-Jacobi equation of the second order:

\begin{equation}\label{eq:HJc2}
\frac{\partial S}{\partial t}+\frac{1}{2m}(\triangledown S)^{2}+V(\textbf{x},t) -i \frac{\hbar}{2m} \vartriangle S =0\text{ \ \ \ \ \ \ \ }\forall \left(
\textbf{x},t\right) \in \mathbb{R} ^{n}\times \mathbb{R}^{+}
\end{equation}
\begin{equation}\label{eq:condinitialHJc2}
S(\textbf{x},0)=S_{0}(\textbf{x})\text{ \ \ \
\ \ }\forall \textbf{x}\in \mathbb{R} ^{n}
\end{equation}
where the $S(\textbf{x},t)$ phase is the complex (Hamilton-Jacobi) action of the quantum particle. To generalize, we extend the $\textbf{x}$ positions to complex numbers $z\in \mathbb{C}$ in dimension 1 ($n=1)$ and $\textbf{z} \in \mathbb{C}^{2}$ in dimension 2 ($n=2)$.
The second generalization is to replace the point particle by an extended particle. For simplicity, this particle is represented by several interacting points. 
This discrete representation allows us to obtain Heisenberg's equations, to make the spin emerge in dimension 2, but also to define a simple generalization of the principle of least action for such an extended particle.

\section{An extended particle model in dimension one}

In dimension 1 ($n=1$), the idea is to represent a particle extended by two interacting points (indexed by $j = 1$ or $2$). We study their evolution from the two vectors $u^1= + 1$ and $ u^2= -1$, and from the permutation $s$ which passes from one vector to the other: $s u^1 =u^2$ and $s u^2 =u^1$; and $s^2 u^1 =u^1$ and $s^2 u^2 =u^2$.

Then, for any given (very small) time step $\varepsilon >0$, the evolution of these two points, at time $t=N \varepsilon $ with $N=2q+r$ ($N,q,r$ integers with $r=0$ or $1$), is defined by the real part of the two discrete processes $z_{\varepsilon }^{j}(t)\in \mathbb{C}$:
\begin{equation}\label{eqprocessus1a}
z_{\varepsilon }^{j}(n \varepsilon )=z_{\varepsilon
}^{j}\left( n\varepsilon -\varepsilon \right)+ v(2q\varepsilon
)\varepsilon +w^{n,j}~~~~with~~~~
z_{\varepsilon }^{j}(0)=z_{0}~~
 for~j=1,2,
\end{equation}
where $w^{n,j}= \gamma(s^{n}u^{j}-s^{n-1}u^{j}) $ is a discrete process of period $2 \varepsilon $, $\gamma= (1+i)\sqrt{\frac{%
\hslash \varepsilon }{2m}}$,
$v(t)$ corresponding to a continuous complex function,
$\hslash  $ is Planck's constant, $m$ the mass of the
particle, and $z_{0}$ is a given vector of
$\mathbb{C}$.

\begin{Remarque}\label{r:interpretationprocessus2a}- The two processes $\textbf{z}_{\varepsilon }^{j}(t)$ resemble Nelson's two stochastic processes (forward and backward) \cite{Nelson1966, Nelson1985} based on Wiener's processes. Indeed, if one poses
\begin{equation}\label{eqprocessus1a2}
E_t [dw^j(t)]=\frac{1}{2}\sum^{2q+2}_{n=2q+1} w^{n,j}~~~~et~~~~E_t [dw^i(t)dw^j(t)]=\frac{1}{2}\sum^{2q+2}_{n=2q+1} w^{n,i}w^{n,j} 
\end{equation}
 at the second order, we find the same white noise properties  
\begin{equation}\label{eqprocessus1a3}
E_t [dw^j(t)]=0~~~~~~~~et~~~~~~~~E_t [dw^i(t)dw^j(t)]= 4 \gamma^2 \delta_i^j,
\end{equation}
but unlike Nelson's processes they are
deterministic. However, these processes seem to be random because, at instant t, the rest modulo
$2$ of the number $n=\frac{t}{\varepsilon}$ is a pseudo-random number.
\end{Remarque}

The average process $\widetilde{z}_{\varepsilon}(t)= \frac{1}{2}(z_{\varepsilon}^{1}(t)+z_{\varepsilon}^{2}(t))$ verifies the discrete system:
\begin{equation}\label{eqprocessus1ba}
\widetilde{z}_{\varepsilon}(n\varepsilon
)=\widetilde{z}_{\varepsilon}(n\varepsilon- \varepsilon)+ v(2q\varepsilon)\varepsilon,~~~~~
\widetilde{z_{\varepsilon}}(0)=z_{0}
\end{equation}
with $t=n\varepsilon=(2q+r)\varepsilon$
and, when $\varepsilon$ tends $0$, $\widetilde{z}_{\varepsilon}(t)$ converges towards $\widetilde{z}(t)$, with the solution to the classical  differential equation:
\begin{equation}
\frac{d\widetilde{z}(t)}{dt}= v(t)~~~~with~~~~
\widetilde{z}(0)=z_{0}
\end{equation} with $v(t)$ defined in equation~(\ref{eqprocessus1a}).
We verify by recurrence that we have at all times
$t= n \varepsilon$:
\begin{equation}\label{eqreccurenceprocessa}
z_{\varepsilon }^{j}(n\varepsilon )=\widetilde{z}_{\varepsilon
}(n\varepsilon )+(1+i) \sqrt{\frac{\hslash \varepsilon }{2m}}\left(
s^{n}u^{j}-u^{j}\right) .
\end{equation}
which leads to $
z_{\varepsilon }^{j}( t)=\widetilde{z}_{\varepsilon }(t )+
O(\sqrt{\varepsilon})$ for all $j$ and all $t = n\varepsilon$.
As $\widetilde{z}_{\varepsilon }(t )=\widetilde{z}(t)+
O(\varepsilon)$ for all $t=n \varepsilon$, we deduce $ z_{\varepsilon
}^{j}( t)=\widetilde{z}(t )+ O(\sqrt{\varepsilon})$. Thus both processes $\widetilde{z}^j_{\varepsilon }(t )$
continuously converges towards
$\widetilde{z}(t)$ when $\varepsilon\rightarrow 0^+$.

As $s^{2}u^{j}=u^{j}$, we deduce from
(\ref{eqreccurenceprocessa}) that
$z_{\varepsilon }^{j}( 2q\varepsilon )=\widetilde{z}_{\varepsilon }(2q
\varepsilon )$ for all $j$. The real part $\widetilde{x}_{\varepsilon }(t) $ of the process $\widetilde{z}_{\varepsilon }(t)$  can be
interpreted as the average position of the particle. The position $x_{\varepsilon }^{j}(t)$ of each point $j$, the real part of the process
$z_{\varepsilon }^{j}(t)$, satisfies the equation:
\begin{equation}\label{eqreccurenceprocessreela}
x_{\varepsilon }^{j}(N\varepsilon )=\widetilde{x}_{\varepsilon
}(n\varepsilon )+ \sqrt{\frac{\hslash \varepsilon }{2m}}\left(
s^{n}u^{j}-u^{j}\right).
\end{equation}

\begin{figure}[H]
\begin{center}
\includegraphics[width=0.6\linewidth]{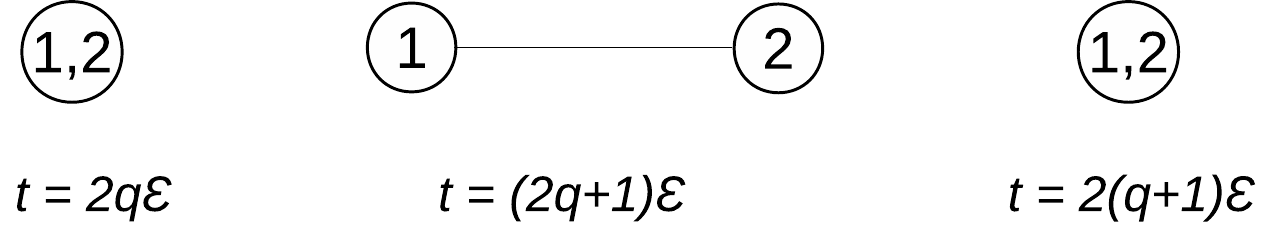}
\caption{\label{fig:sixprocessus1}Evolution of the two points of the extended particle over a period of time 
$2\varepsilon$ from left to right (punctual at $2q\varepsilon$, extending to $(2q+1)\varepsilon$, then punctual again at $2(q+1)\varepsilon$).}
\end{center}
\end{figure}

This equation yields the evolution of the two points of the extended particle with respect to its center-of-mass. The evolution of this extended particle over a period of $2\varepsilon$ is visualized in Figure~\ref{fig:sixprocessus1}. We can consider that the two points $x_{\varepsilon }^{j}(t)$
of the particle define the structure of an elastic cord. The movement of its two
points corresponds to the vibration of the string.
At the instant $t=2 q \varepsilon $, the two ends are joined together and the length of the string at that instant is therefore zero. At the instant $t=2(q+1)\varepsilon$,
it has an extension.
Let us show that these positions verify Heisenberg's uncertainty relations. The standard deviation of the position $\langle\Delta x\rangle$ defined by $\langle\Delta x\rangle^2 =\frac{1}{4}\sum^{2q+2}_{n=2q+1} \sum^2_{j=1} (x_{\varepsilon }^{j}(n\varepsilon )-\widetilde{x}_{\varepsilon}(n\varepsilon ))^2$ is equal to $ \frac{\hbar
\varepsilon}{2m}  $. The standard deviation of the momentum $p_\varepsilon^j(n\varepsilon)=m\frac{x_{\varepsilon }^{j}(n\varepsilon )-\widetilde{x}_{\varepsilon}(n\varepsilon )}{\varepsilon}$ defined by $\langle\Delta p\rangle^2 =\frac{1}{4}\sum^{2q+2}_{n=2q+1} \sum^2_{j=1} (p_{\varepsilon }^{j}(n\varepsilon )-\widetilde{p}_{\varepsilon}(n\varepsilon ))^2$ is equal to $ \frac{\hbar m}{2 \varepsilon}  $, which yields 
$$\langle \Delta x\rangle \cdot \langle \Delta
p\rangle = \frac{\hslash }{2}.$$

\section{An extended particle model in dimension 2}

In an orthonormal marker of space $ \mathbb{R}^2$, we consider the four vertices $ u^1=\binom{1}
{1}$, $ u^2=\binom{1}
{-1}$, $ u^3=\binom{-1}
{-1}$ et $ u^4=\binom{-1}
{1}$ of the unit square. There are two circular permutations of these four vertices, one $s^+$ clockwise, the other $s^-$ counter-clockwise. We have for each of these two permutations $s\in S=\{s^-, s^+\}$ and for all $u^j$ ($j=1...4$), $s^4 u^j= u^j$. 

We consider an extended particle represented by four points. For any step of time $\varepsilon >0 $ and at each of the two permutations $s\in S$, the evolution of these four points, at the time $t=N \varepsilon $ with $N=4q+r$ ($N,q,r$ integers with $0\leq r\leq 3 $), is defined by the real part of the 4 following discrete processes $\textbf{z}_{\varepsilon }^{j}(t)\in \mathbb{C}^{2}$:
\begin{equation}\label{eqprocessus1}
\textbf{z}_{\varepsilon }^{j}(N \varepsilon )=\textbf{z}_{\varepsilon
}^{j}\left( n\varepsilon -\varepsilon \right)+ \textbf{v}(4q\varepsilon
)\varepsilon +\gamma(s^{n}u^{j}-s^{n-1}u^{j})~~\text{with}~~
\textbf{z}_{\varepsilon }^{j}(0)=\textbf{z}_{0},~
 \forall j
\end{equation}
where  $\gamma= (1+i)\sqrt{\frac{%
\hslash \varepsilon }{4m}}$,
$\textbf{v}(t)$ corresponds to a continuous complex function, $\hslash $\ is the Planck constant, $m$ the mass of the particle, and $\textbf{z}_{0}$ is a given vector of
$\mathbb{C}^{2}$.

\begin{figure}[H]
\begin{center}
\includegraphics[width=0.9\linewidth]{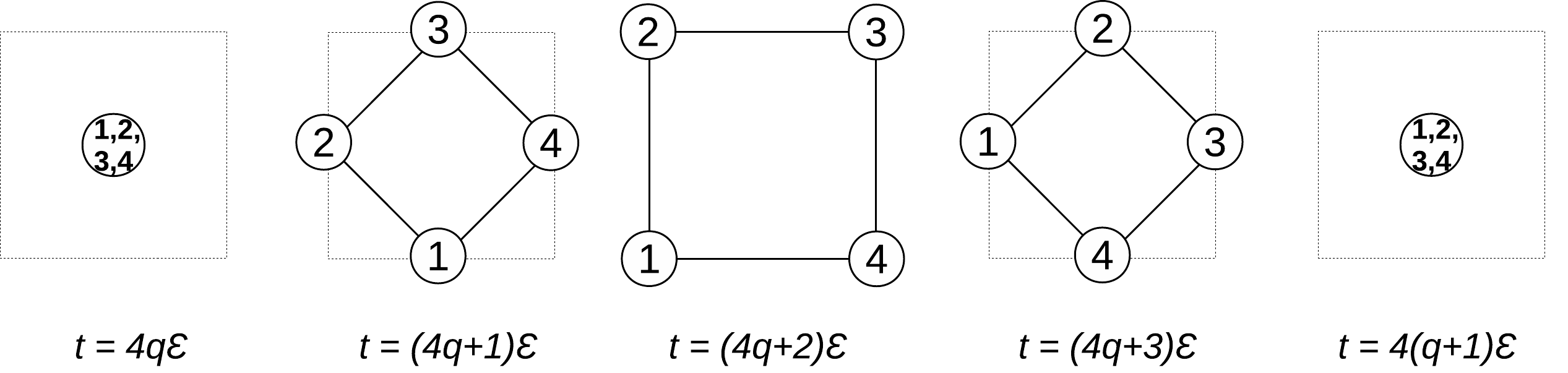}
\caption{\label{fig:sixprocessus}Evolution of the four points of the extended particle over a period of time 
4 $\varepsilon$ from left to right (punctual at $4q\varepsilon$, in extension at $( 4q+1)\varepsilon$ and $(4q+2)\varepsilon$, then in contraction at $(4q+3)\varepsilon$ and $4(q+1)\varepsilon$).}
\end{center}
\end{figure}

Let $\widetilde{\textbf{z}}_{\varepsilon }(t)$ be the solution in $\mathbb{C}
^{2} $ of the discrete system of the instant $%
t=N\varepsilon$ with $N=4q+r$ ($N,q$ and $r$
 integers and $0\leq r\leq 3$) by the equation:
\begin{equation}\label{eqprocessus1b}
\widetilde{\textbf{z}}_{\varepsilon }(N \varepsilon
)=\widetilde{\textbf{z}} _{\varepsilon }(N \varepsilon- \varepsilon
)+ \textbf{v}(4q\varepsilon )\varepsilon,~~~~~
\widetilde{\textbf{z}_{\varepsilon }}(0)=\textbf{z}_{0}.
\end{equation}
We then verify by recurrence that we have at any time
$t= N \varepsilon$:
\begin{equation}\label{eqreccurenceprocess}
\textbf{z}_{\varepsilon }^{j}(N\varepsilon )=\widetilde{\textbf{z}}_{\varepsilon
}(N\varepsilon )+(1+i) \sqrt{\frac{\hslash \varepsilon }{4m}}\left(
s^{n}u^{j}-u^{j}\right) .
\end{equation}
As $s^{4}u^{j}=u^{j}$, we deduce from
(\ref{eqreccurenceprocess}) that $\textbf{z}_{\varepsilon }^{j}( 4q\varepsilon )=\widetilde{\textbf{z}}_{\varepsilon }(4q
\varepsilon )$ for all $j$. As $\sum^{j=4}_{j=1}s^n u^j =0$, we deduce from (\ref{eqprocessus1}) that the process $\widetilde{\textbf{z}}_{\varepsilon }(t)$ is the average of the four processes $\textbf{z}_{\varepsilon }^{j}(t)$. Its real part $\widetilde{\textbf{x}}_{\varepsilon }(t) $  can be
interpreted as the average position of the particle. The position $\textbf{x}_{\varepsilon }^{j}(t)$ of each point $j$, real part of the process
$\textbf{z}_{\varepsilon }^{j}(t)$, satisfies the equation:
\begin{equation}\label{eqz}
\textbf{x}_{\varepsilon }^{j}(n\varepsilon )=\widetilde{\textbf{x}}_{\varepsilon
}(n\varepsilon )+ \sqrt{\frac{\hslash \varepsilon }{4m}}\left(
s^{n}u^{j}-u^{j}\right).
\end{equation}
This equation demonstrates the evolution of the four points of the extended particle in relation to its center of mass. The evolution of this extended particle over a period of $4\varepsilon $ is visualized in figure
\ref{fig:sixprocessus}.
We can consider that the four points $\textbf{x}_{\varepsilon }^{j}(t)$
of the particle define the structure of a closed string. The movement of its four
points corresponds to the vibration of the string.
At the moment $t=4 q \varepsilon $, the four
points are located in the center of a square and the length of the string at that moment is therefore zero. At the moments $t \neq 4 q \varepsilon $,
it takes an extension. At times $(4q+1)\varepsilon $ and
$(4 q+3)\varepsilon $, the four points are located in the center of the  sides of the
square. At the moment $(4q+2)\varepsilon $, the four
points are located on all four
corners of the square.  Furthermore,
this interpretation suggests a kind of creative process
between moments $t=4 \varepsilon $ and $(4q+1)\varepsilon $ followed by
a process of annihilation between moments $(4 q+3)\varepsilon$
and $(4q+4)\varepsilon $.

The equation (\ref{eqreccurenceprocess}) yields 
$\textbf{z}_{\varepsilon }^{j}( t)=\widetilde{\textbf{z}}_{\varepsilon }(t )+ O(\sqrt{\varepsilon})$ for all $j$ and all $t = N\varepsilon$. Let $\widetilde{\textbf{z}}(t)$ be the solution to the differential equation:
\begin{equation}
\frac{d\widetilde{\textbf{z}}(t)}{dt}= \textbf{v}(t)~~~~with~~~~
\widetilde{\textbf{z}}(0)=\textbf{z}_{0}.
\end{equation}
Since $\textbf{v}(t)$ is continuously differentiable, we
have $\widetilde{\textbf{z}}_{\varepsilon }(t )=\widetilde{\textbf{z}}(t)+
O(\varepsilon)$ for any $t=N \varepsilon$, and therefore $\textbf{z}_{\varepsilon }^{j}( t)=\widetilde{\textbf{z}}(t )+ O(\sqrt{\varepsilon})$. We deduce that each process $\widetilde{\textbf{z}}^j_{\varepsilon }(t )$
continuously converges towards
$\widetilde{\textbf{z}}(t)$ when $\varepsilon\to 0^+$.

\begin{Remarque}\label{r:interpretationFeynman}- 
In \cite{Feynman1965}, Feynmann and Hibbs
show that the "important paths" of quantum mechanics, although continuous, are very irregular and nowhere differentiable. They admit an average speed\\$
\displaystyle\lim_{\Delta t\to 0^+}\langle \frac{x_{k+1}-x_k}{\Delta
t} \rangle =v$,
but not a root mean square speed because $
\displaystyle\langle (\frac{x_{k+1}-x_k}{\Delta t})^2 \rangle =\frac{i \hbar}{m
\Delta t}$.
The four processes $\textbf{x}_{\varepsilon }^{j}(t)$ verify the same properties as Feynmann's paths, being increasingly
irregular and non-differentiated when $\varepsilon
\rightarrow 0^{+}$; however, the value of $\varepsilon=\Delta t$, although very
small, remains finite. 
\end{Remarque}

\section{Emergence of spin and Heisenberg's uncertainty relations}

An extended particle was associated with four points and a four-point variation cycle during the period $T =4 \varepsilon$. It is assumed that the properties of such a particle correspond to the average of the properties of the four points taken over the four instants of the period.  

We therefore define the mean angular momentum of the extended particle verifying
the equation (\ref{eqreccurenceprocess}) by:
\begin{equation*}
\sigma =E_{n,j}(\sigma^j_n)=\frac{1}{16}
\sum_{n=4q}^{4q+3}\sum_{j=1}^{4} \sigma^j_n
\end{equation*}
with
\begin{equation*}
\sigma^j_n=r^j_n \wedge p^j_n,~~~~r^j_n=\textbf{x}^j_\varepsilon
(n\varepsilon)= \widetilde{r}_n + \sqrt{\frac{\hbar
\varepsilon}{4m}}(s^n u^j - u^j)
\end{equation*}
and
\begin{equation*}
p^j_n=m \frac{r^j_{n+1}-r^j_n}{\varepsilon}= m
\frac{\widetilde{r}_{n+1}-\widetilde{r}_n}{\varepsilon}+
\sqrt{\frac{\hbar m}{4 \varepsilon}}(s^{n+1} u^j - s^n u^j).
\end{equation*}
Using the relationship $\sum_{j=1}^{4} s^n u^j=0$ for all n,
we obtain $\sigma =\wedge m\widetilde{r}\wedge m\widetilde{v}+\frac{1}{16}\hslash
\sum_{j=1}^4 (u^j \wedge s u^j)$
with $ \widetilde{r}=\frac{1}{4} \sum_{n=4q}^{4q+3} \widetilde{r}_n$
and $\widetilde{v}=\widetilde{v}(4q\varepsilon)$.
For $s=s^+$, we have
\begin{equation*}
\sigma \equiv \sigma _{z}=m\left( x\widetilde{v}
_{y}-y\widetilde{v}_{x}\right) -\frac{\hslash }{2}.
\end{equation*}
\begin{Proposition}\label{th:emergencespin}- For any $\varepsilon > $0, the extended particle corresponding to the real part of the process
(\ref{eqprocessus1}), admits an angular moment
intrinsic average $s_{z}=-\frac{\hslash }{2}$ for permutation $s^+$ and $s_{z}=+ \frac{\hslash }{2}$ for permutation $s^-$. 
\end{Proposition}

Let $\widetilde{x}_{\varepsilon }(n\varepsilon )$ be the position
average on the $x$ axis of the particle  at the moment
$n\varepsilon $ (with $n=4q+r$) and $m\widetilde{v}( 4q \varepsilon ) $ its average impulse. The calculation of
standard deviations $\triangle x$\ and $\Delta p_{x}$ of the position
and the impulse on the $x$ axis is obtained from the following equations:
\begin{equation*}
\langle \Delta x\rangle^2=\frac{1}{16}\sum_{n=4q}^{4q+3}\sum_{j=1}^{4}\left( r^j_n- \widetilde{r}_n\right) _{x}^{2}
\end{equation*}
\begin{equation*}
\langle \Delta p_{x}\rangle^2=\frac{1}{16}\sum_{n=4q}^{4q+3}\sum_{j=1}^{4}\left(
p_n^j- \widetilde{p}_n\right) _{x}^{2}
\end{equation*}
with $\widetilde{p}_n= m
\frac{\widetilde{r}_{n+1}-\widetilde{r}_n}{\varepsilon}$. We find $\langle \Delta x\rangle=\frac{\hbar
\varepsilon}{2m}$ et $\langle \Delta p_{x}\rangle= \frac{\hbar m}{ \varepsilon}$.
\begin{Proposition}\label{th:inegaliteheisenberg}- For any $\varepsilon >$0
and for any $s$, the extended particle corresponding to the real part of the process
(\ref{eqprocessus1}) satisfies Heisenberg's uncertainty relations:
\begin{equation}
\langle \Delta x\rangle \cdot \langle \Delta
p_{x}\rangle = \frac{\hslash }{2}.
\end{equation}
\end{Proposition}

Let $f$ be a class C$^{2}$ application from $\mathbb{C}^{2}\times \mathbb{R}$ in $\mathbb{C}$.
We consider the \textit{Dynkin complex} operator, introduced by Nottale \cite{Nottale1993} under the name
of "quantum covariant derivatives", by:
\begin{equation}
D=\frac{\partial }{\partial t}+ \textbf{v} \cdot \triangledown -i\frac{%
\hslash }{2m}\triangle.
\end{equation}

\begin{Proposition}\label{th:lemme} - For all $\varepsilon >0$ and for all $s $, the process
$Y_{\varepsilon }(t)$ defined by:
\begin{equation}
Y_{\varepsilon }\left( t\right) =Ef(\textbf{z}_{\varepsilon }^{j}\left( t\right) ,t)=\frac{1%
}{4}\ \sum_{j}\left( f\left( \textbf{z}_{\varepsilon }^{j}\left( t\right)
,t\right) \right)
\end{equation}
with $\textbf{z}_{\varepsilon }^{j}\left( t\right) $ built on
the equation (\ref{eqprocessus1}),
satisfies for all $t= 4q\varepsilon$ (q integer):
\begin{equation}\label{eqdevsecondordre}
Y_{\varepsilon }(t) -Y_{\varepsilon }\left(t - \varepsilon  \right) =Df\left( \widetilde{\textbf{z}}\left(t\right) ,t \right) \varepsilon +0\left( \varepsilon^2 \right).
\end{equation}
\end{Proposition}
\textbf{Proof}: First of all, we have $Y_{\varepsilon }\left(
4q\varepsilon \right) =f\left( \widetilde{\textbf{z}}\left( 4q\varepsilon \right)
,4q\varepsilon \right) $. In
using (\ref{eqreccurenceprocess}) and taking into account $
\sum^{j=4}_{j=1}s^{n}u^{j}=0 $, then we have for every $t=N \varepsilon$
\begin{equation*}
Ef(\textbf{z}_{\varepsilon }^{j}\left( t\right) ,t)=f(\widetilde{\textbf{z}}_{\varepsilon
}(t),t)+
\frac{i\hslash \varepsilon}{4m} E\left\{ \sum_{k,l}\frac{\partial ^{2}f(\widetilde{\textbf{z}}_{\varepsilon
}(t),t)}{%
\partial x_{k}\partial x_{l}}\left( s^{n}u^{j}-u^{j}\right) _{k}\left(
s^{n}u^{j}-u^{j}\right) _{l}\right\} +0 \left( \varepsilon^2 \right).
\end{equation*}
For $n=4q-1$, $E (
s^{n}u^{j}-u^{j})_{k}(s^{n}u^{j}-u^{j})_{l}=\frac{4+4}{4}\delta_{kl}$
and the calculation of the last term of $Ef(\textbf{z}_{\varepsilon }^{j}\left(
t\right) ,t)$ yields $\frac{i\hslash \varepsilon}{4m}
 2 \Delta f$. We deduce:
\begin{equation*}
Y_{\varepsilon }\left( 4q \varepsilon-\varepsilon \right)
=f(\widetilde{\textbf{z}}_{\varepsilon
}(4q \varepsilon-\varepsilon ),4q\varepsilon -\varepsilon )+i\frac{\hslash \varepsilon}{2m} \Delta f(%
\widetilde{\textbf{z}}_{\varepsilon }(4q \varepsilon -\varepsilon ),4q\varepsilon -\varepsilon )+0
\left( \varepsilon^2 \right)
\end{equation*}
hence equation (\ref{eqdevsecondordre}) developed thanks to (\ref{eqprocessus1b} )
$f(\widetilde{\textbf{z}}_{\varepsilon }(4q \varepsilon-\varepsilon ),4q \varepsilon-\varepsilon )$
in order one.$\Box $

\begin{Remarque}\label{r:prop3dim1}- We obtain the analog of proposition 3 for the one-dimensional particle model constructed on the equations (\ref{eqprocessus1a}) of section 3. 
\end{Remarque}

\section{The complex Hamilton-Jacobi equation of the second order}

We now show that the evolution corresponding to the processes $\textbf{z}_{\varepsilon }^{j}(t) $ defined by the equations (\ref{eqprocessus1}), is governed by a complex second-order Hamilton-Jacobi equation. To do so, we use an analytical mechanic 
complex and a generalized principle of least action. The analytical mechanism
complex is a generalization of classical analytical mechanics, with objects having a complex position
$\textbf{z}(t)\in \mathbb{C}^{2}$, a complex speed $\textbf{v}\left(
t\right) \in \mathbb{C}^{2}$, and using the minimum of one
complex function and the minmost-complex analysis we
introduced in
\cite{Gondran2001a,Gondran2001c,Gondran2003}
and whose principle we recall in Definition 1 and Definition 2.
\begin{Definition}- For a complex function $f\left( \textbf{z}\right) =f\left( \textbf{x} +i \textbf{y}\right)
$ from $\mathbb{C}^{n}$ to $\mathbb{C}$ written under the
form $f\left( \textbf{z}\right) =P\left( \textbf{x},\textbf{y}\right) +iQ\left( \textbf{x},\textbf{y}\right)
$, the complex minimum, if it exists, is defined as follows
$\min \left\{ f\left( \textbf{z}\right) /\textbf{z}\in \mathbb{C}^{n}\right\}
=f\left( \textbf{z}_{0}\right) $ where $\left( \textbf{x}_{0},\textbf{y}_{0}\right) $ must
to be a point of $P\left( \textbf{x},\textbf{y}\right)$:
$P\left( \textbf{x}_{0},\textbf{y}\right) \leq P\left( \textbf{x}_{0},\textbf{y}_{0}\right) \leq
P\left( \textbf{x},\textbf{y}_{0}\right)$ $\forall
(\textbf{x},\textbf{y})\in R^{n}\times R^{n}$.
\end{Definition}

We say that a complex function $f\left( \textbf{z}\right) $ is
(strictly) \textit{convex} if $P\left( \textbf{x},\textbf{y}\right) $\ is
(strictly) convex in $\textbf{x}$\ and (strictly) concave in $\textbf{y}$. We
verify that, if $f\left( \textbf{z}\right) $ is in addition holomorphic, the minimum condition becomes $\nabla f(\textbf{z}) = $0.

Then, from a classical Lagrange function $L(x,\dot{x}
,t)$, an analytical function in $x$ and $\dot{x}$, we define by analytical prolongation
the complex Lagrange function $L(\textbf{z},\textbf{v} ,t)$.

\begin{Definition}- For any complex function that is strictly convex, we
associate its complex Fenchel-Legendre transform $\textbf{p}\in \mathbb{C}%
^{n}\longmapsto \widehat{f}\left( \textbf{p}\right) \in \mathbb{C}$
ended by:
\begin{equation*}
\widehat{f}\left( \textbf{p}\right) =\underset{\textbf{z}\in \mathbb{C}^{n}}{\mathit{\max }}%
\left( \textbf{p}.\textbf{z}-f\left(\textbf{z}\right)\right)
\end{equation*}
\end{Definition}

We can then define a complex Hamilton-Jacobi action for the extended particle corresponding to the processes (\ref{eqprocessus1}). As for the classical case, we give ourselves, at the initial instant, a complex Hamilton-Jacobi action $\mathcal{S}^{0}\left( \textbf{z}\right) $, and a holomorphic function of $\mathbb{C}^{2}$ in $\mathbb{C}$. As announced in Note 1, the complex Hamilton-Jacobi action associated with process (\ref{eqprocessus1}) is obtained from a generalization of the optimality equation (\ref{eq:defactionlocaleb}).

\begin{Definition}:\textbf{Principle of least generalized action} - The Hamilton-Jacobi action is defined as complex
$\mathcal{S}_{\varepsilon }(\textbf{z},t)$ at times $ t=4q\varepsilon $ ($q \geq $1) by the following optimality equation:
\begin{equation}\label{eq:principemoindreactiongeneral}
\mathcal{S}_{\varepsilon }(\textbf{z} ,t)=\min_{\textbf{v}%
\left( t\right) }\frac{1}{4}\sum_{j}\left\{
\mathcal{S}_{\varepsilon
}(\textbf{z}- \textbf{v}(t)\varepsilon -\gamma(s^4 u^j -s^3u^j),t-\varepsilon )+L(\textbf{z}, \textbf{v}(t),t)\varepsilon \right\}
\end{equation}
where the minimun is taken in the sense of the minimum complex on the
possible complex speeds $\textbf{v}\left( t\right)$, and for $t=0$ by the initial condition:
\begin{equation*}
\mathcal{S}_{\varepsilon }\left( \textbf{z},0\right) =\mathcal{S}^{0}\left(
\textbf{z}\right) \text{ \ \ \ \ \ \ }\forall \textbf{z}\in \mathbb{C}^{2}.
\end{equation*}
\end{Definition}

At time $t= 4q \varepsilon$, we have $
\widetilde{\textbf{z}_{\varepsilon
}}(t )= \textbf{z}_{\varepsilon }^{j}(t)= \textbf{z}_{\varepsilon }^{j}(t-\varepsilon)+ \textbf{v}(t
)\varepsilon +\gamma(s^{4}u^{j}-s^{3}u^{j})$ for all j. We obtain the optimality equation
(\ref{eq:principemoindreactiongeneral}) from the optimality equation (\ref{eq:defactionlocaleb}) by identifying $\textbf{x}$ to $\textbf{z} $ and $\textbf{u}(t)\varepsilon $ to the various $\textbf{v}(t)
)\varepsilon +\gamma(s^{4}u^{j}-s^{3}u^{j}) $.
Equation (\ref{eq:principemoindreactiongeneral}) can thus be
interpreted as a new principle of lesser action adapted to the
processes defined by (\ref{eqprocessus1}).
In this case, the decision on the command is made only at the
instants $t=4q\varepsilon $, that is to say, at the instants
corresponding to annihilation-creation.

\begin{Theoreme}\label{th:eqschrodingeraction}- If a complex process satisfies the principle of
generalized action (\ref{eq:principemoindreactiongeneral}) and admits
$L(x,\dot{x},t)= \frac{1}{2}m\dot{x}^{2}-V\left( x\right) $ like
a Lagrangian, then the complex action verify
the Hamilton-Jacobi equation complex of the second
order:
\begin{equation}\label{eqHJsecondordrecomplexe1}
\frac{\partial \mathcal{S}}{\partial t}+\frac{1}{2}\left(
\triangledown \mathcal{S}\right) ^{2}+V\left( \textbf{z}\right)
-i\frac{\hslash }{2m}\triangle
\mathcal{S}=0\text{ \ \ \ \ \ \ }\forall \left(\textbf{z},t\right) \in \mathbb{C}%
^{2}\times \mathbb{R}^{+}
\end{equation}
\begin{equation}\label{eqHJsecondordrecomplexe2}
\mathcal{S}\left( \textbf{z},0\right) =\mathcal{S}^{0}\left( \textbf{z}\right)
\text{ \ \ \ \ \ \ }\forall \textbf{z}\in \mathbb{C}^{2}
\end{equation}
and the complex speed is given for each (\textbf{z},t) by the speed field:
\begin{equation}\label{vitesseimagine}
\textbf{v}\left(\textbf{z}, t\right) =\frac{\triangledown \mathcal{S}\left(
\textbf{z},t\right) }{m}.\text{ \ \ \ \ \ \ }\forall \left(\textbf{z},t\right) \in \mathbb{C}%
^{2}\times \mathbb{R}^{+}
\end{equation}
\end{Theoreme}

\textbf{Proof}: Formal proof is made here only in
assuming that $ \mathcal{S}_{\varepsilon }(\textbf{z},t)$ is a 
very regular function in $\varepsilon$ and that it is holomorphic in $\textbf{z}$ and
derivable in t. Thanks to proposition \ref{th:lemme}, by taking f=$S_\varepsilon$, one obtains:
\begin{equation*}
\frac{1}{4}\sum_{j}\left\{ \mathcal{S}_{\varepsilon
}(\textbf{z}- \textbf{v}(t)\varepsilon -\gamma(s^4 u^j- s^3u^j)),t- \varepsilon
)\right\} =\mathcal{S}_{\varepsilon }(\textbf{z},t)-D\mathcal{S}_{\varepsilon
}(\textbf{z},t)\varepsilon +0(\varepsilon^2).
\end{equation*}
We deduce at the point $\left( \textbf{z},t\right) $ the following equation:

\begin{equation}\label{eqtransformedefenchelb}
\frac{\partial \mathcal{S}_{\varepsilon }}{\partial t}=\underset{\textbf{v}}{%
\text{\textit{min}}}\left( L(\textbf{z}, \textbf{v}{\LARGE
,}t)- \textbf{v}\cdot
\triangledown \mathcal{S}_{\varepsilon }+i\frac{\hslash }{2m}\triangle \mathcal{%
S}_{\varepsilon }+0\left( \varepsilon \right) \right)
\end{equation}
hence equation
(\ref{eqHJsecondordrecomplexe1}) is obtained by making $\varepsilon $ tend towards $0
^{+} $, and then taking the transform from
the Fenchel
complex of $L(\textbf{z},\textbf{v},t)$. 
As $L(\textbf{z},\textbf{v},t)=\frac{1}{2}m \textbf{v}^{2}-V\left(
\textbf{z},t\right) $, the minimum of (\ref{eqtransformedefenchelb}) is obtained
when $m \textbf{v}-\triangledown \mathcal{S}_{\varepsilon }=0,$ which leads to (\ref{vitesseimagine}). $\Box $

\section{Schrödinger's equation "pilots" the center of mass}

By defining as wave function $\Psi
=e^{i\frac{\mathcal{S}}{\hslash }}$ and taking the restriction
of
(\ref{eqHJsecondordrecomplexe1}) (\ref{eqHJsecondordrecomplexe2})and (\ref{vitesseimagine})
to the real part $\textbf{x}$ of $\textbf{z}$,
theorem \ref{th:eqschrodingeraction} allows to deduce:

\begin{Theoreme}\label{th:schrodingerreel}- If a complex process verifies the principle of generalized least action 
(\ref{eq:principemoindreactiongeneral}) and admits $L(\textbf{x},\dot{\textbf{x}},t)=
\frac{1}{2}m\dot{\textbf{x}}^{2}-V\left( \textbf{x},t\right) $ as a Lagrangian, then
its wave function satisfies 
Schr\"{o}dinger's equation:
\begin{equation}\label{eqschrodingera}
i\hslash \frac{\partial \Psi }{\partial t}=\mathcal{-}\frac{\hslash ^{2}}{2m}%
\triangle \Psi +V(\textbf{x},t)\Psi \qquad \forall (\textbf{x},t)\in
\mathbb{R}^{2}\times \mathbb{R}^{+}
\end{equation}
\begin{equation}\label{eqschrodingerb}
\Psi (\textbf{x},0)=\Psi ^{0}(\textbf{x})\qquad \forall \textbf{x}\in \mathbb{R}^{2}.
\end{equation}
Moreover, the $\widetilde{\textbf{x}}(t)$ center of mass of the particle follows a trajectory whose velocity is given at each point ($\textbf{x}$,t) by the velocity field
proposed by de Broglie~\cite{deBroglie1927} and Bohm~\cite{Bohm1952}:
\begin{equation}\label{eqBBtrajectoire}
\textbf{v}(\textbf{x},t)= \frac{\triangledown S (\textbf{x},t)}{m}.
\end{equation}
where $S(\textbf{x},t)$ is the phase of the wave function $\Psi(\textbf{x},t) $ written in the semi-classical representation  $\Psi(\textbf{x},t) = \sqrt{\rho (\textbf{x},t)}e^{i\frac{S(\textbf{x},t)}{\hbar}}$.
\end{Theoreme}

\textbf{Proof}: As we have assumed $\mathcal{S}(\textbf{z},t) $ to be holomorphic in $\textbf{z}$, equations (\ref{eqschrodingera}) and (\ref{eqschrodingerb}), where $\Psi(\textbf{x},t)$ is the restriction of $\Psi(\textbf{x},t)$ to the real $\textbf{x}$ part of $\textbf{z}$, can be deduced from equations (\ref{eqHJsecondordrecomplexe1}) and (\ref{eqHJsecondordrecomplexe2}).
Since $S(\textbf{x},t)$ is the real part of the complex action
\begin{equation}\label{actioncomplexe}
\mathcal{S}(\textbf{x},t)=S(\textbf{x},t)-i \frac{\hslash}{2} log \rho(\textbf{x},t),
\end{equation}
we deduce (\ref{eqBBtrajectoire}) from equation (\ref{vitesseimagine}) and the center of gravity of the particle
 $\widetilde{\textbf{x}}(t)$ satisfies the differential equation:
\begin{equation}\label{eqBBtrajectoireb}
\frac{d\widetilde{{\large \textbf{x}}}(t)}{dt}=\frac{\nabla S (\textbf{x},t)}{m}\mid_{\textbf{x}= \widetilde{ \textbf{x}}(t)},\text{ \
\ \ \ \ \ \ \ \ }\widetilde{{\large \textbf{x}}}(0)={\large \textbf{x}}_{0}.
\end{equation}

To specify the model, one must choose the $ \varepsilon$ time step. To define it, we make two hypotheses: Inside the extended particle, the passage from one state to the next is done at the speed of light and the distances correspond to Compton's wavelength $ \lambda _{C}$ as seems to be shown with Foldy-Wouthuysen's transformation \cite{Foldy1950}. So we shall assume:
\begin{equation}
4\varepsilon =T \simeq \frac{\lambda _{C}}{c}=\frac{h}{mc^{2}}.
\end{equation}

\begin{Remarque}\label{r:th2et3dim1}- As in Remark 3, the analogy of theorems 2 and 3 is obtained for the one-dimensional particle model constructed on the equations (\ref{eqprocessus1a}) of Section 3. In this case, we obtain:
\begin{equation}
2\varepsilon =T \simeq \frac{\lambda _{C}}{c}=\frac{h}{mc^{2}}.
\end{equation}
\end{Remarque}

\begin{Remarque}\label{r:discret}- How do we interpret our discrete particle models? Are there fictitious particles represented by the different points? Or is there a continuous model of the particle whose different points correspond to a sampling whose frequency is twice the maximum frequency of the particle? Thanks to Nyquist-Shannon's sampling theorem, it is not necessary to choose between these two hypotheses. 
One can also wonder whether the $\textbf{x}^j_\varepsilon (n \varepsilon) $ process of equations (\ref{eqreccurenceprocessreela}) and (\ref{eqz}), composed of a classical $\widetilde {\textbf{x}}_\varepsilon (n \varepsilon) $ part and a small periodic oscillatory part, could be connected to Schrödinger's Zitterbewegung~\cite{Schrodinger1930}.
\end{Remarque}

\section{Conclusion}
By taking up the idea of Nelson's stochastic processes with pseudo-random and non-point processes, we have shown that in dimension 2 there is an extended quantum particle model that admits a spin and verifies Heisenberg's equations. Thanks to a generalization of the optimality equation from the principle of least action to a complex action for an extended particle, we have associated it with a wave function that verifies the Schrödinger equation. The center of gravity of this quantum particle model then follows the de Broglie-Bohm trajectories generalizing to quantum mechanics the case of the center of gravity of a classical particle which is driven by the classical Hamilton-Jacobi action.  

This particle model corresponds to a deepening of the wave-particle duality in quantum mechanics: an extended particle with a small extension vibrates like a small string. The equation for the evolution of its center of gravity verifies a time-dependent Schrödinger equation, i.e. the equation for a wave that spreads over time.

We thus find through a different method the interpretation of the double scale-theory that we have just defined \cite{Gondran2021}. Indeed, by taking up the idea of Louis de Broglie's double-solution theory, we show that the complete wave function of a quantum system, such as an atom or a molecule, is the product of two wave functions: an external wave function for the evolution of its center of mass and an internal function for the evolution of its internal variables in the reference frame of the center of mass. These two wave functions do not correspond to the same scale and have different meanings and interpretations. The external wave function represents the macroscopic view of the quantum system: it "drives" the center of mass and spin and corresponds to the de Broglie-Bohm "pilot wave". The internal wave function represents the microscopic view of the quantum system: our interpretation is the one proposed by Schrödinger at the 1927 Solvay congress: the particles are expanded and the square of the modulus of the (internal) wave function of an electron corresponds to the density of its charge in space. This double solution (wave and soliton) depending on the scales clearly explains the wave-corpuscle duality.  It clarifies the debates on the interpretation of quantum mechanics, which did not differentiate between external and internal wave functions.

By placing ourselves in Clifford's algebra $Cl_3$, it should be possible to extend this model to dimension 3 with a spinner verifying Pauli's equation and a spin orienting itself during the measurement operation as in Stern and Gerlach's experiment
\cite{Gondran2016a}.

The hypothetical model we have just presented is only one possible model that can have many variants.
It is also compatible with Lorentz and Poincaré's electron model \cite{Poincare1906} as well as with Dirac's extensible model \cite{Dirac1962}.

\renewcommand\refname{References}
\bibliographystyle{abbrv}
\bibliography{biblio_mq}

\begin{thebibliography}{10}

\bibitem{Bacciagaluppi2009}
G.~Bacciagaluppi and A.~Valentini.
\newblock {\em Quantum Theory at the Crossroads: Reconsidering the 1927 Solvay
  Conference}.
\newblock Cambridge University Press, 2009.
\newblock arXiv:quant-ph/0609184.

\bibitem{Bohm1952}
D.~Bohm.
\newblock {A Suggested interpretation of the quantum theory in terms of hidden
  variables.}
\newblock {\em Physical Review}, 85:166--193, 1952.

\bibitem{deBroglie1927}
L.~de~Broglie.
\newblock {La m{\'e}canique ondulatoire et la structure atomique de la
  mati{\`e}re et du rayonnement}.
\newblock {\em Jounal de physique}, 8:225--241, 1927.
\newblock English translation in~\cite{Bacciagaluppi2009}.

\bibitem{Dirac1962}
P.~Diarc.
\newblock An extensible model of the electron.
\newblock {\em Proc. R. Soc. Lond. A}, 1332(268):57--47, 1962.

\bibitem{Feynman1965}
R.~Feynman and A.~Hibbs.
\newblock {\em {Quantum Mechanics and Paths Integrals}}.
\newblock McGraw-Hill, 1965.

\bibitem{Foldy1950}
L.~L. Foldy and S.~A. Wouthuysen.
\newblock On the dirac theory of spin 1/2 particles and its non-relativistic
  limit.
\newblock {\em Phys. Rev.}, 78:29--36, Apr 1950.

\bibitem{Gondran2001c}
M.~Gondran.
\newblock {Calcul des variations complexe et solutions explicites
  d'{\'e}quations d'Hamilton-Jacobi complexes}.
\newblock {\em C. R. Acad. Sci. - Series I}, 332(7):677--680, 2001.

\bibitem{Gondran2001a}
M.~Gondran.
\newblock {Processus complexe stochastique non standard en m{\'e}canique}.
\newblock {\em C. R. Acad. Sci. Paris - Series I}, 333(6):592--598, 2001.

\bibitem{Gondran2004a}
M.~Gondran.
\newblock {Schr{\"o}dinger Equation and Minplus Complex Analysis}.
\newblock {\em Russian Journal of Mathematical Physics}, 11(2):130--139, 2004.

\bibitem{Gondran2016a}
M.~Gondran and A.~Gondran.
\newblock {Replacing the Singlet Spinor of the EPR-B Experiment in the
  Configuration Space with Two Single-Particle Spinors in Physical Space}.
\newblock {\em Foundations of Physics}, 46(9):1109--1126, 2016.
\newblock arXiv:1504.04227.

\bibitem{Gondran2021}
M.~Gondran, A.~Gondran, and C.~No{\^u}s.
\newblock {The Two-scale Interpretation: de Broglie and Schr{\"o}dinger's
  External and Internal Wave Functions}.
\newblock {\em Annales de la Fondation Louis de Broglie}, 46(1):87--126, 2021.

\bibitem{Gondran2003}
M.~Gondran and R.~Saade~Hoblos.
\newblock Complex calculus of variations.
\newblock {\em Kybernetika}, 39(2):[249]--263, 2003.

\bibitem{Nelson1966}
E.~Nelson.
\newblock {Derivation of the Schr\"odinger Equation from Newtonian Mechanics}.
\newblock {\em Phys. Rev.}, 150:1079--1085, Oct 1966.

\bibitem{Nelson1967}
E.~Nelson.
\newblock {\em {Dynamical Theories of Brownian Motion}}.
\newblock Princeton University Press, Princeton, 1967.

\bibitem{Nelson1985}
E.~Nelson.
\newblock {\em {Quantum fluctuations}}.
\newblock Princeton University Press, Princeton, 1985.

\bibitem{Nelson2012}
E.~Nelson.
\newblock Review of stochastic mechanics.
\newblock {\em Journal of Physics: Conference Series}, 361(1):012011, may 2012.

\bibitem{Nottale1993}
L.~Nottale.
\newblock {\em Fractal Space-Time and Microphysics : Towards a Theory of Scale
  Relativity}.
\newblock World Scientific (Singapore, New Jersey, London), 1993.

\bibitem{Poincare1906}
H.~Poincar{\'e}.
\newblock {Sur la dynamique de l'{\'e}lectron}.
\newblock {\em Rend. Circ. Matem. Palermo}, 21:129--175, Dec. 1906.

\bibitem{Schrodinger1930}
E.~Schr{\"o}dinger.
\newblock {\"U}ber die kr{\"a}ftefreie bewegung in der relativistischen
  quantenmechanik.
\newblock volume~24, pages 418--428. Sitz. Preuss. Akad. Wiss. Phys. Math. Kl.,
  1930.

\end{thebibliography}

\end{document}